\documentclass[12pt]{article}

\usepackage{epsfig}

\def\beq   {\begin{equation}}
\def\eeq   {\end{equation}}
\def\beqd  {\begin{displaymath}}
\def\eeqd  {\end{displaymath}}
\def\beqaa {\begin{eqnarray}}
\def\eeqaa {\end{eqnarray}}

\newcommand{\bea}{\begin{eqnarray}}
\newcommand{\eea}{\end{eqnarray}}

\def\noi {\noindent}

\def\ti  {\tilde}

\def\st{\ifmmode{\tilde{t}} \else{$\tilde{t}$} \fi}
\def\sb{\ifmmode{\tilde{b}} \else{$\tilde{b}$} \fi}
\def\sq{\ifmmode{\tilde{q}} \else{$\tilde{q}$} \fi}
\def\sg{\ifmmode{\tilde{g}} \else{$\tilde{g}$} \fi}

\def\a   {\alpha}
\def\b   {\beta}
\def\t   {\theta}

\def\sz{\ifmmode{\tilde{\chi}^0} \else{$\tilde{\chi}^0$} \fi}
\def\sw{\ifmmode{\tilde{\chi}} \else{$\tilde{\chi}$} \fi}

\def\sQ{{\tilde Q}}
\def\sU{{\tilde U}}
\def\sD{{\tilde D}}
\def\msQ{{M_{\tilde Q}}}
\def\msU{{M_{\tilde U}}}
\def\msD{{M_{\tilde D}}}

\begin{document}

\pagestyle{empty}

\vspace*{-1cm} 
\begin{flushright}
  HEPHY-PUB 725/99 \\
  TGU-25 \\
  TU-583 \\
  hep-ph/9912463
\end{flushright}

\vspace*{1.4cm}

\begin{center}

{\Large {\bf
Improved SUSY QCD corrections to Higgs boson decays 
into quarks and squarks} 
}\\

\vspace{10mm}

{\large 
H.~Eberl$^a$, K.~Hidaka$^b$, S.~Kraml$^a$, 
W.~Majerotto$^a$, and Y.~Yamada$^c$}

\vspace{6mm}
\begin{tabular}{l}
$^a${\it Institut f\"ur Hochenergiephysik der \"Osterreichischen Akademie
der Wissenschaften,}\\
\hphantom{$^a$}{\it A--1050 Vienna, Austria}\\
$^b${\it Department of Physics, Tokyo Gakugei University, Koganei,
Tokyo 184--8501, Japan}\\
$^c${\it Department of Physics, Tohoku University, Sendai 980--8578, Japan}
\end{tabular}

\end{center}

\vspace{3cm}
\vfill

\begin{abstract}
We improve the calculation of the supersymmetric 
${\cal O}(\alpha_s)$ QCD corrections to the decays of  
Higgs bosons into quarks and squarks 
in the Minimal Supersymmetric Standard Model. 
In the on--shell renormalization scheme 
these corrections can be very large, which makes the 
perturbative expansion unreliable. This is especially serious 
for decays into bottom quarks and squarks for large $\tan\beta$. 
Their corrected widths can even become negative. 
We show that this problem can be solved by a careful choice of the 
tree--level Higgs boson couplings to quarks and squarks, 
in terms of the QCD and SUSY QCD running quark masses, 
running trilinear couplings $A_q$, and on--shell left--right mixing 
angles of squarks. 
We also present numerical results for the corrected partial 
decay widths for the large $\tan\beta$ case. 
\end{abstract}

\newpage
\pagestyle{plain}
\setcounter{page}{2}

\section{Introduction} 

The Minimal Supersymmetric Standard Model (MSSM) \cite{mssm} is 
a very promising extension of the Standard Model. 
This model has five physical Higgs scalars, 
$h^0$, $H^0$, $A^0$, and $H^\pm$. Studying their properties is 
very important for probing the nature of the breaking of the electroweak 
symmetry and supersymmetry. 

The Higgs scalars have large couplings to quarks and 
scalar quarks (squarks) of the third generation \cite{gh}. 
Therefore, in typical cases the decays to top and bottom quarks have large 
branching ratios and are often the main modes, see e.\,g. \cite{review}. 
Decays to top and bottom squarks can be 
also dominant \cite{hsqtree,hsqsugra} if they are kinematically 
allowed and the left--right mixing parameters of the squarks are large. 
Studying these decays is therefore very important for the discovery 
and the detailed study of the Higgs bosons in the MSSM. 

The decays of Higgs bosons into quarks and squarks receive 
large radiative corrections by QCD interactions. 
In the MSSM, one--loop SUSY QCD corrections to the decay 
widths have been given 
analytically, both for quark modes \cite{dabel,hbb,htb,htb2} and 
squark modes \cite{hsq,hsq2}. 
However, when the lowest order widths are given in terms of 
the on--shell parameters for quarks and squarks, 
the ${\cal O}(\a_s)$ correction terms are often comparable to or 
even larger than the lowest order ones, especially for decays into 
bottom quark and squark. 
Such large corrections make the perturbation calculation of the 
decay widths quite unreliable. 
When the correction term is negative, 
the corrected width can even become negative, 
which clearly makes no sense. This ``negative width'' problem has
also been observed in other processes of squarks \cite{sqhx}. 

The couplings of Higgs bosons to quarks and squarks depend on the 
quark masses $m_q$. 
For the decays into quarks, it is well known \cite{BL,hdecay} that 
the large part of the gluon loop correction 
can be absorbed by giving the tree--level Higgs--quark couplings 
in terms of the QCD running quark masses $m_q(Q)_{\rm SM}$, where $Q$ 
is at the scale of the parent Higgs boson mass. 
The convergence of the perturbation series is greatly improved 
by this replacement. 
However, the gluino loop correction to the decays into 
bottom quarks can also be very large for large $\tan\beta$, the 
ratio of the vacuum expectation values of two Higgs fields. 
It has been pointed out \cite{dmb} that 
the main source of this correction is the counterterm for $m_b$. 
Its contribution to the decay widths into bottom quarks was included 
in ref.~\cite{hbbnew} by effective Higgs--bottom couplings 
with one free parameter. 
However, no improvement of the large gluino loop corrections 
has been presented along this line. 
Moreover, the Higgs boson couplings to squark are functions 
of $m_q$, Higgs--squark trilinear couplings $A_q$, 
and squark left--right mixing angles $\theta_{\sq}$.  
The corresponding improvement of large corrections to squark modes 
is therefore more complicated and has not been discussed so far. 

In this paper we improve the one--loop SUSY QCD corrected widths 
of the MSSM Higgs boson decays into quarks and squarks. 
We concentrate on the decays into bottom quarks and squarks, 
where the large QCD corrections are most serious in a phenomenological 
study. 
The essential point of the improvement is to define appropriate 
tree--level couplings of the Higgs bosons to quarks and 
squarks, in terms of the running quark masses $m_q(Q)$ both in 
non--SUSY and SUSY QCD, running Higgs--squark trilinear 
couplings $A_q(Q)$, 
and on--shell left--right mixing angle $\theta_{\sq}$ of squarks. 
Such a treatment of the Higgs boson couplings to quarks and squarks 
will also be useful in studying radiative corrections to other 
processes of Higgs bosons and squarks. 

This paper is organized as follows. In section 2, we review the 
improvement of the large gluon loop corrections to the decays into 
quarks by using running quark masses. In section 3, we discuss the 
origin of the large gluino loop corrections to 
the decays into bottom quarks in the large $\tan\beta$ case,
and work out a suitable method of improvement. 
The corresponding improvement of the squark modes is discussed in 
section 4. Section 5 shows numerical results of the partial decay 
widths of Higgs bosons into bottom quarks and squarks. 
Section 6 is devoted to conclusions.
Throughout this paper we adopt notations and conventions of \cite{hsq}. 

\section{Gluon loop corrections to Higgs decays to quarks}

We first review the method to improve the calculation of the gluon 
loop corrections to Higgs decays to quarks $\phi\rightarrow q\bar{q}$ 
($\phi=h^0,H^0,A^0$) by using QCD running quark masses and 
renormalization group equation, following \cite{BL,hdecay}. 
A similar discussion holds for the decay $H^{\pm}\rightarrow q\bar{q}'$. 

When the Yukawa couplings in the lowest--order decay widths of a 
Higgs boson $\phi$ are given in terms of the pole masses $M_q$ of the 
quarks, the one--loop QCD corrections to the inclusive decay widths 
(including real gluon radiation) have 
large ${\cal O}(\alpha_s\ln(m_\phi/M_q))$ terms from gluon loops. 
The explicit form of the corrected width is \cite{BL,hdecay}, 
in the limit of $M_q\ll m_\phi$, 
\beq  \label{e1}
\Gamma^{\rm corr}=\Gamma^0_{\rm OS}
\left[ 1+\frac{\alpha_s}{\pi}\left( -2\ln\frac{m_\phi^2}{M_q^2}+3 
\right) \right]. 
\eeq
The correction can be very large and causes bad convergence of 
perturbation series. This problem is especially crucial 
for the decays to $b$ quarks. 
In calculations using dimensional regularization 
with renormalization scale $Q=m_\phi$, 
the above correction comes from the counterterm $\delta m_q$ 
for the Higgs--quark Yukawa coupling $h_q\propto m_q$. 
It is well known \cite{BL,hdecay} that 
this correction can be absorbed into the tree--level widths 
by using the (non--SUSY) QCD running quark masses 
$m_q(Q\sim m_\phi)_{\rm SM}$ for the tree--level Yukawa couplings. 
We can also resum large higher order corrections by using 
renormalization group equations for quark masses. 

The one--loop relation between $M_q$ and the 
$\overline{\rm MS}$ (dimensional regularization with modified minimal 
subtraction) running mass $m_q(Q)_{\rm SM}$ in the non--SUSY QCD is 
\beq 
m_q(Q)_{\rm SM}=M_q+\delta m_q^{(g)}(Q). \label{e2}
\eeq
The counterterm $\delta m_q^{(g)}(Q)$ from the gluon loop is
at one--loop level 
\beq
\delta m_q^{(g)}(Q) = 
-\frac{\alpha_s(Q)}{\pi}M_q\left( \ln\frac{Q^2}{M_q^2}+\frac{4}{3}
\right) . \label{e3}
\eeq
One can see that the large logarithmic part of Eq.~(\ref{e1}) is 
cancelled when $m_q(m_\phi)_{\rm SM}$ is used in the tree--level width. 

Next we rewrite Eq.~(\ref{e2}) as 
\beq
m_q(Q)_{\rm SM}=\left(\frac{m_q(Q)_{\rm SM}}{m_q(M_q)_{\rm SM}}\right)
\, m_q(M_q)_{\rm SM}\, .
\label{e4}
\eeq
In Eq.~(\ref{e4}) a large logarithmic correction 
appears only in the ratio $m_q(Q)_{\rm SM}/m_q(M_q)_{\rm SM}$. 
We can then resum large higher--order corrections of 
${\cal O}(\alpha_s^n\ln^n(m_\phi/M_q))$ in Eq.~(\ref{e4}) 
by using the renormalization group equations (RGE) 
for $m_q(Q)_{\rm SM}$. 
The one--loop running $\alpha_s(Q)$ is given by 
\beq
\alpha_s(Q) = \frac{12\pi}{(33-2\,n_f)\ln (Q^2/\Lambda^2_{n_f})}\,. \label{e6}
\eeq
The solutions of the two--loop RGEs \cite{hdecay} are
\begin{equation}   
\alpha_s^{(2)}(Q) = \frac{12\pi}{(33-2\,n_f)\ln \frac{Q^2}
{\Lambda^2_{n_f}}}     
\left( 1-\frac{6(153-19\,n_f)}{(33-2\,n_f)^2}                                
\frac{\ln\ln\frac{Q^2}{\Lambda^2_{n_f}}}{\ln\frac{Q^2}{\Lambda^2_{n_f}}}
\right)\, ,                                                           
\label{eq:as2loop} 
\end{equation}
and for the running quark mass at $M_q$
\begin{equation} 
m_q(M_q)_{\rm SM}=M_q\left[1+\frac{4}{3}\frac{\alpha_s^{(2)}(M_q)}{\pi}   
+K_q\left(\frac{\alpha_s^{(2)}(M_q)}{\pi}\right)^2\right]^{-1}\, ,        
\end{equation}                                                    
with $K_t\sim 10.9$ and $K_b\sim 12.4$ and 
$n_f=5$ or 6 for $M_b < Q \le M_t$ or $Q > M_t$, respectively.
Using the functions
\begin{eqnarray}
c_5(x)&=& \left(\frac{23}{6}x\right)^{\frac{12}{23}}(1+1.175x)\;\;
(M_b < Q \le M_t)\,,\\
c_6(x)&=& \left(\frac{7}{2}x\right)^{\frac{4}{7}}(1+1.398x)\;\;
(Q > M_t)\,,
\end{eqnarray}
the solution for the ratio 
$\left(m_q(Q)_{\rm SM}/m_q(M_q)_{\rm SM}\right)$ is
\begin{eqnarray}
  \label{eq:RGEsol1}
M_b < Q \le M_t: &\quad&
\frac{c_5(\alpha_s^{(2)}(Q)/\pi)}{c_5(\alpha_s^{(2)}(M_q)/\pi)}\, ,\\
Q > M_t: &\quad& 
\frac{c_6(\alpha_s^{(2)}(Q)/\pi)}{c_6(\alpha_s^{(2)}(M_t)/\pi)}\,
\frac{c_5(\alpha_s^{(2)}(M_t)/\pi)}{c_5(\alpha_s^{(2)}(M_q)/\pi)}\, .
\label{eq:RGEsol2}
\end{eqnarray}
For the numerical calculations in section~5 
we take $Q=m_\phi$. For the calculation of $m_q(Q)_{\rm SM}$ we use
eq.~({\ref{e4}}) and eqs.~(\ref{eq:as2loop}) -- (\ref{eq:RGEsol2}). 
For all other
calculations eq.~({\ref{e6}}) is used.

\section{Gluino corrections to quark Yukawa couplings}

\subsection{Correction to \boldmath $m_b$}

The gluino loop corrections to the decay widths of 
$\phi\rightarrow b\bar{b}$ \cite{hbb} and 
$H^+\rightarrow t\bar{b}$ \cite{htb,htb2} 
can be very large for the cases with large $\tan\beta = v_2/v_1$, 
large $m_{\sg}$, and large $|\mu|$. 
Their main part comes from the counterterm 
$\delta m_b^{(\sg)}$ by squark--gluino loops which has 
an enhancing coefficient $\tan\beta$. 
In contrast to the case of gluon loops, however, 
these corrections are not caused by a large logarithm. 
Rather, as we will see now, it originates from the effective 
$\bar{b}bH_2$ coupling which is generated by the soft SUSY breaking 
in the loops. 

Here we follow the discussion in ref.~\cite{dmb,hbbnew,chankowski}. 
For illustration, let us consider the case where squarks are 
much heavier than Higgs bosons. 
In this case, the interactions between Higgs scalars and quarks 
are properly described by the effective two Higgs doublets 
theory, after integrating out squarks and gluino. 
The effective interactions of Higgs scalars and b 
which are proportional to $m_b$ are written as 
\beq
{\cal L}_{\rm int}^{\rm eff}= 
-h_b \bar{b}_R(b_L H_1^0 - t_L H_1^-) 
-h_b\Delta_b \bar{b}_R(b_L H_2^{0*} + t_L H_2^-) 
+({\rm h.c.}).  \label{e7}
\eeq

At the tree--level, the $\bar{b}bH_2$ coupling is forbidden by SUSY, 
namely $\Delta_b=0$. However, this coupling is generated by loop corrections 
with soft SUSY breaking. 
Indeed, an explicit calculation of the squark--gluino loops 
for zero external momenta gives the result 
\beq
\Delta_b = -\frac{2\alpha_s}{3\pi}m_{\sg}\mu I(m_{\sg}, 
M_{\tilde{Q}}, M_{\tilde{D}}), \label{e8}
\eeq
where 
\beqaa
I(m_1,m_2,m_3)&=&C_0(0,0,0;m_1,m_2,m_3) \nonumber \\
&=& 
\frac{m_1^2m_2^2\ln(m_1^2/m_2^2)+m_2^2m_3^2\ln(m_2^2/m_3^2)+
m_3^2m_1^2\ln(m_3^2/m_1^2)}
{(m_1^2-m_2^2)(m_2^2-m_3^2)(m_3^2-m_1^2)}. \label{e9}
\eeqaa
$C_0$ is the standard three--point function \cite{pave} in the convention 
of \cite{denner}. 
We see that $\Delta_b=0$ for the exact SUSY case $m_{\sg}=0$. 
As long as ($m_{\sg}$, $|\mu|$) are not much larger than 
$M_{\tilde{Q},\tilde{D}}$, $\Delta_b$ itself is smaller than 
unity and therefore does not destroy the validity of 
perturbative calculation. 
Nevertheless, as shown below, $\Delta_b$ is responsible for the very 
large gluino correction to the $\phi\rightarrow b\bar{b}$ 
and $H^+\rightarrow t\bar{b}$ decays. 

We then evaluate the couplings of Higgs bosons to $\bar{b}b$ 
in the mass eigenstate basis. For reference, we list 
the basic relations between the gauge and mass bases \cite{gh}: 
\beq
\sqrt{2}{\rm Re}\,
\left( \begin{array}{c} H_1^0 \\ H_2^0 \end{array} \right) = 
\left( \begin{array}{rr} \cos\alpha & -\sin\alpha \\ 
                         \sin\alpha & \cos\alpha 
\end{array} \right) 
\left( \begin{array}{c} H^0 \\ h^0 \end{array} \right) 
+\left( \begin{array}{c} v_1 \\ v_2 \end{array} \right)\, ,
\eeq
\beq
\sqrt{2}{\rm Im}\,
\left( \begin{array}{c} H_1^0 \\ H_2^0 \end{array} \right) = 
\left( \begin{array}{rr} -\cos\beta & \sin\beta \\ \sin\beta & \cos\beta 
\end{array} \right) 
\left( \begin{array}{c} G^0 \\ A^0 \end{array} \right) \, ,
\eeq
\beq
\left( \begin{array}{c} H^+_1 \\ H^+_2 \end{array} \right) = 
\left( \begin{array}{rr} -\cos\beta & \sin\beta \\ \sin\beta & \cos\beta 
\end{array} \right) 
\left( \begin{array}{c} G^+ \\ H^+ \end{array} \right) \, .
\eeq
$G^{0,\pm}$ are the would--be Nambu--Goldstone bosons and irrelevant here. 
In the large $m_A$ limit, 
$\cos\alpha\rightarrow \sin\beta$ and $\sin\alpha\rightarrow -\cos\beta$ 
follows. 

In the mass basis, Eq.~(\ref{e7}) becomes ($\bar{v}^2\equiv v_1^2+v_2^2$) 
\beqaa
{\cal L}_{\rm int}^{\rm eff}&=& 
-\frac{h_b\bar{v}}{\sqrt{2}}[\cos\beta+\Delta_b \sin\beta]\,
\bar{b}b \nonumber \\
&&-\frac{h_b}{\sqrt{2}}[\cos\alpha+\Delta_b \sin\alpha]\,H^0\bar{b}b 
\nonumber \\
&&+\frac{h_b}{\sqrt{2}}[\sin\alpha-\Delta_b \cos\alpha]\,h^0\bar{b}b 
\nonumber \\
&&+\frac{ih_b}{\sqrt{2}}[\sin\beta-\Delta_b \cos\beta]\,A^0\bar{b}\gamma_5b 
\nonumber \\
&&+h_b[\sin\beta-\Delta_b \cos\beta]\,H^-\bar{b}_R t_L +({\rm h.c.}).
\label{e13}
\eeqaa
The first term gives the (non--SUSY) QCD running mass $m_b|_{\rm SM}$ 
at scale $Q$, 
\beq
m_b(Q)_{\rm SM}=
\frac{h_b\bar{v}}{\sqrt{2}}[\cos\beta+\Delta_b \sin\beta] . \label{e14}
\eeq
The difference from the SUSY QCD running mass 
$m_b(Q)_{\rm MSSM}=h_b\bar{v}\cos\beta/\sqrt{2}$ in the 
$\overline{\rm DR}$ scheme (dimensional reduction with modified minimal 
subtraction) is due to 
the $\Delta_b$ term. For $\tan\beta \gg 1$, 
the effect of $\Delta_b$ on $m_b$ is enhanced by a large vacuum expectation 
value of $H_2$. As a result, the relative correction 
by squark--gluino loops 
$(m_b(Q)_{\rm SM}-m_b(Q)_{\rm MSSM})/m_b(Q)_{\rm MSSM}$ has a 
factor $\tan\beta$ and can become very large, even ${\cal O}(1)$. 
This property has been pointed out in \cite{dmb}, 
mainly as a weak--scale threshold correction to the bottom--tau Yukawa 
coupling unification. 

In Eq.~(\ref{e13}) the contributions of $\Delta_b$ to 
the Higgs--bottom couplings take forms different from those to $m_b$. 
When the tree--level couplings are given in terms of $m_b(Q)_{\rm SM}$, 
the corrections by $\Delta_b$ can be very enhanced for $\tan\beta\gg 1$, 
as pointed out in \cite{chankowski,hbbnew}. 
For example, $A^0\bar{b}b$ coupling is expressed as 
\beqaa
a^b &=& 
\frac{i\,h_b}{\sqrt{2}}[\sin\beta-\Delta_b \cos\beta] \nonumber \\
&=&
\frac{i\,m_b|_{\rm SM}}{\bar{v}}\tan\beta\left[ 
1-\frac{1}{\sin\beta\cos\beta}\frac{\Delta_b}{1+\Delta_b\tan\beta}
\right] . \label{e15}
\eeqaa
Similarly, the corrections to the couplings $H^0\bar{b}b$, 
$h^0\bar{b}b$, and $H^+\bar{t}b$ can also be very large. 
We see that $\Delta_b$ is the main source 
of the large squark--gluino loop corrections to decay widths of 
$(H^0,A^0)\rightarrow b\bar{b}$ \cite{hbb} and 
$H^+\rightarrow t\bar{b}$ \cite{htb} in the 
on--shell renormalization scheme. 
Since the largeness of the gluino correction comes from the property 
that the $H_2\bar{b}b$ coupling is forbidden at tree--level, 
higher order corrections are not expected to be larger than the one--loop 
corrections. 

Here we have two comments. First, similar to the $H_2\bar{b}b$ coupling, 
the squark--gluino loops 
also generate an effective $H_1\bar{t}t$ coupling with a coefficient 
$\Delta_t h_t$.
As consequence, for large $\tan\beta$ large gluino loop corrections
of ${\cal O}(\alpha_s\Delta_t\tan\beta)$ to the
decay widths of $(A^0,H^0)\rightarrow t\bar{t}$ appear
in the vertex corrections \cite{hbb}.
However, their widths and branching ratios decrease
as $\tan\beta$ increases. The contribution of $\Delta_t$ to
the width of $H^+\rightarrow t\bar{b}$ is sufficiently small.
Therefore, the phenomenological importance of the $\Delta_t$
correction is smaller than that of $\Delta_b$.
Second, the effective coupling $\Delta_b$ can be also
generated from other loop corrections such as the higgsino loops. 
This effect has also been 
discussed \cite{dmb,chankowski,htbyukawa,hbbnew}, but 
it is beyond the scope of this paper.

\subsection{Method of improvement} \label{sec3.2}

It is clear from Eqs.~(\ref{e13})--(\ref{e15}) that 
the QCD expansion of the Higgs decay widths to bottom quarks 
around the tree--level ones in terms of $M_b$ or $m_b(m_\phi)_{\rm SM}$ 
may cause bad convergence. As in the case of the gluon loops, we can 
improve the QCD perturbative expansion by changing 
the definition of the tree--level coupling of bottom quarks and Higgs bosons. 
Ref.~\cite{hbbnew} calculated the decay widths into 
quarks by using effective couplings in Eqs.~(\ref{e13})--(\ref{e15})
and added the gluon loop corrections.
However, numerically, $\Delta_b$ was not calculated from the SUSY
parameters.
Here we take a different approach, which is more suitable for our 
purpose to improve the SUSY QCD correction. 

The simplest case is the decay $A^0\rightarrow \bar{b}b$. 
The coupling $a^b$, Eq.~(\ref{e15}),
is expressed in terms of the SUSY QCD 
running mass $m_b(Q)_{\rm MSSM}$ at $Q=m_A$ as 
\beq
a^b = \frac{i\,m_b(Q)_{\rm MSSM}}{\bar{v}}\tan\beta [1-\Delta_b \cot\beta].
\label{e16}
\eeq
In contrast to Eq.~(\ref{e15}), the correction is now very small. 
This is due to the property that, for $\tan\beta\gg 1$, 
$A^0$ is almost ${\rm Im}\,H_1^0$ while $\Delta_b$ is the 
coupling to $H_2$. We therefore expect that the SUSY QCD running mass 
$m_b(Q=m_A)_{\rm MSSM}$ is an appropriate parameter for the tree--level 
$A^0\rightarrow\bar{b}b$ decay. Since $H^\pm$ is also almost 
$H_1^\pm$ for large $\tan\beta$, $m_b(m_{H^\pm})_{\rm MSSM}$ is 
appropriate for the $H^+\rightarrow t\bar{b}$ decay. 

The case of ($H^0$, $h^0$) decays needs a special treatment. 
The couplings $s^b_H$ for $H^0\bar{b}b$ and 
$s^b_h$ for $h^0\bar{b}b$ in terms of $m_b(Q)_{\rm MSSM}$ 
receive relative corrections $\Delta_b\tan\alpha$ and 
$-\Delta_b\cot\alpha$, respectively. 
One of them can be very large. 
For example, for very large $m_A$, $s^b_h$ becomes 
\beq
s^b_h \rightarrow -\frac{m_b(Q)_{\rm MSSM}}{\bar{v}}[1+\Delta_b \tan\beta]
=\frac{-m_b(Q)_{\rm SM}}{\bar{v}}. \label{e17}
\eeq
This is the SM decoupling limit for $h^0$.
In this case $m_b(Q)_{\rm SM}$ is more appropriate 
than $m_b(Q)_{\rm MSSM}$ for the tree--level coupling $s^b_h$. 

To find a better way, we make a change of the basis of 
the Higgs doublets to 
\beq
H_{\rm sm}\equiv \cos\beta H_1^c + \sin\beta H_2, \;\;\;
H_{\rm ex}\equiv -\sin\beta H_1^c + \cos\beta H_2,  \label{e18}
\eeq
where $H_1^c\equiv(-H_1^+,H_1^{0*})$. 
Only $H_{\rm sm}$ has a vacuum expectation value. $H_{\rm sm}$ 
can be regarded as the ``true Higgs'', while 
$H_{\rm ex}$ is the ``extra scalar doublet''. ($A^0$, $H^{\pm}$) 
are included in $H_{\rm ex}$. $H^0$ and $h^0$ have components of 
both $H_{\rm sm}$ and $H_{\rm ex}$. 

The couplings of $H_{\rm sm}$ and $H_{\rm ex}$ are obtained from 
Eq.~(\ref{e7}). 
One can see that the $H_{\rm sm}\bar{b}b$ coupling takes the form 
$h_bc_{\beta}(1+\Delta_b\tan\beta)$, which is properly parametrized by 
$m_b(Q)_{\rm SM}$. On the other hand, the $H_{\rm ex}\bar{b}b$ coupling is 
$h_bs_{\beta}(1-\Delta_b\cot\beta)$, which is properly parametrized by 
$m_b(Q)_{\rm MSSM}$. Using the relation between ($H^0$, $h^0$) 
and ($H_{\rm sm}$, $H_{\rm ex}$), 
\beq
\sqrt{2}\,{\rm Re}
\left( \begin{array}{c} H^0_{\rm sm} \\ H^0_{\rm ex} \end{array} \right) = 
\left( \begin{array}{rr} \cos(\alpha-\beta) & -\sin(\alpha-\beta) \\ 
\sin(\alpha-\beta) & \cos(\alpha-\beta) 
\end{array} \right) 
\left( \begin{array}{c} H^0 \\ h^0 \end{array} \right) , \label{e19}
\eeq
the appropriate choices for the tree--level couplings are 
\beqaa
s^b_H &=& -\cos(\alpha-\beta)\,\frac{m_b(Q)_{\rm SM}}{\bar{v}} 
+ \sin(\alpha-\beta)\tan\beta \,\frac{m_b(Q)_{\rm MSSM}}{\bar{v}}, 
\label{e20}\\
s^b_h &=& \sin(\alpha-\beta)\,\frac{m_b(Q)_{\rm SM}}{\bar{v}}
+ \cos(\alpha-\beta)\tan\beta \,\frac{m_b(Q)_{\rm MSSM}}{\bar{v}}\,.
\label{e21}
\eeqaa
Eq.~(\ref{e17}) is reproduced in the large $m_A$ limit, using 
$\cos(\alpha-\beta)\rightarrow 0$. 

In the numerical calculation, we obtain $m_b(Q)_{\rm MSSM}$ from 
$m_b(Q)_{\rm SM}$ in Eq.~(\ref{e4}) by using 
\beq 
m_b(Q)_{\rm MSSM}=m_b(Q)_{\rm SM}+\delta m_q^{(\sg)}(Q), \label{e22}
\eeq
with the full one--loop contribution of gluino loops and the conversion term 
between the $\overline{\rm MS}$ and $\overline{\rm DR}$ schemes 
\beqaa
\delta m_q^{(\sg)} &=& - \frac{\alpha_s}{3\pi} \left[
M_q(B_1(M_q^2, m_{\sg}^2, m_{\sq_1}^2)
+B_1(M_q^2, m_{\sg}^2, m_{\sq_2}^2)+1) \right. \nonumber\\
&& \left. +m_{\sg} \sin 2\theta_{\sq} (B_0(M_q^2, m_{\sg}^2, m_{\sq_1}^2)
-B_0 (M_q^2, m_{\sg}^2, m_{\sq_2}^2) ) \right] , \label{e23}
\eeqaa
instead of the zero momentum approximation Eqs.~(\ref{e8}), 
(\ref{e9}), and (\ref{e15}). 

The formulae of the effective couplings in this section were obtained
in the limit of $m_A\ll m_{\sq}$.
Nevertheless, we apply these formulae in section~5
to the case of $m_{\sq}<m_A$. This is justified because for the decay
modes discussed in section~5 the modification of
$\Delta_b$ and $\Delta_t$ by large external
momentum does not affect the effective Higgs--quark couplings
significantly.
The situation is different for
the decays to $t\bar{t}$.
We therefore leave such decays for future works.

\section{Higgs--squark couplings} \label{sec4}

We next consider the SUSY QCD corrections to the Higgs boson 
decays to squark pairs. 
As in the case of decays to quarks, an appropriate choice of 
the tree--level couplings of Higgs bosons and squarks is essential
for improving the QCD perturbation calculation. 

We first review the on--shell renormalization of the squark sector, 
following \cite{hsq}. 
For given values of $\mu$ and $\tan\beta$
the masses and mixings of squarks in a generation 
are fixed by five independent parameters, 
in addition to the masses of the quark partners. We can use, 
as such parameters, 
the pole masses ($m_{\st_1}$, $m_{\st_2}$, $m_{\sb_1}$, $m_{\sb_2}$) 
and the on--shell left--right mixing angles ($\theta_{\st}$, $\theta_{\sb}$) 
which are independent of the $\overline{\rm DR}$ scale $Q$, 
with one constraint by SU(2) gauge symmetry. 
The on--shell mixing angle $\theta_{\sq}$ is defined by specifying 
the counterterm 
$\delta\theta_{\sq}=\t_{\sq}(\overline{\rm DR})-\t_{\sq}(\mbox{on--shell})$. 
Various definitions have been proposed in previous works on the 
squark interactions \cite{eesqsq,beenakker3,hsq2,sqyukawa}. 
Here we adopt the definition in \cite{sqyukawa}, 
where $\delta\t_{\sq}$ is fixed to absorb the anti--hermitian 
part of the squark wave--function renormalization:
\beq
\delta\t_{\sq} =
\frac{1}{2}\frac{\mbox{Re}\left\{
\Pi^\sq_{12}(m_{\sq_1}^2)+\Pi^\sq_{12}(m_{\sq_2}^2) \right\}  }
{m_{\sq_1}^2 - m_{\sq_2}^2}  \label{e24}
\eeq
with $\Pi^\sq_{12}(p^2)$ the off--diagonal squark self energy 
in the squark mass basis.
Eq.~(\ref{e24}) is also applicable to loop corrections other than in QCD.  

We can then define the on--shell squark parameters ($M_{\tilde{Q}}$, 
$M_{\tilde{U}}$, $M_{\tilde{D}}$, $A_t$, $A_b$) in terms of the above 
on--shell parameters by using tree--level relations. For example, on--shell 
$A_q$ is given by 
\beq \label{e25} 
M_q A_q = \frac{1}{2}(m_{\sq_1}^2 - m_{\sq_2}^2) \sin 2\t_{\sq} +
M_q \mu \{ \cot \beta, \tan \beta\}\, ,
\eeq
where $\cot\beta\, (\tan\beta)$ is for $\sq = \st\, (\sb)$. 
Also, on--shell $M_{\tilde{Q}}^2(\sq)$ for the $\sq_{1,2}$ sector 
is given by 
\beq   \label{e26}
M_{\ti Q}^2(\sq) = m_{\sq_1}^2 \cos^2\t_{\sq} + 
m_{\sq_2}^2 \sin^2\t_{\sq} 
-m_Z^2 \cos 2\beta (I^{3L}_{q} - e_{q} \sin^2\theta_W) -M_q^2 
\eeq
(with $I^{3L}_q$ the third component of the weak isospin 
and $e_q$ the charge of the quark $q$).
These parameters were used as inputs in \cite{hsq,sqhx}. 
Note that the value of on--shell $M_{\tilde{Q}}^2(\st)$ is different
from that of 
$M_{\tilde{Q}}^2(\sb)$ by finite QCD corrections \cite{hsq,sqhx,cos2b}. 

The QCD correction to the decay width of Higgs bosons into 
squarks consists of four parts: 
vertex correction, squark wave function correction, 
counterterm for the Higgs--squark coupling, and real gluon radiation. 
Since a squark couples to both $H_1$ and $H_2$ at the tree--level, 
the vertex corrections are not expected to give a large correction. 
A large correction to decays into $\sb$ in the on--shell scheme 
comes mainly from the third part \cite{hsq}, the counterterms for the 
Higgs--squark couplings. 
These couplings depend on three parameters which receive 
QCD corrections, $m_q$, $\t_{\sq}$, and $A_q$. In the on--shell 
renormalization, their counterterms can be very large and make 
the perturbation calculation unreliable. 

As in the decays to quarks, we can improve the perturbation 
calculation by 
using SUSY QCD running parameters $m_q(m_\phi)_{\rm MSSM}$ 
and $A_q(m_\phi)$ in the tree--level Higgs--squark 
couplings. However, the mixing angles $\t_{\sq}$ are kept on--shell 
in order to cancel the ${\cal O}(1/(m_{\sq_2}^2-m_{\sq_1}^2))$ term 
in the off--diagonal squark wave function correction, 
which causes a singularity for $m_{\sq_2}\sim m_{\sq_1}$. 
In addition, the counterterm Eq.~(\ref{e24}) largely cancels 
the ${\cal O}(m_q m_{\sg})$ term in the off--diagonal squark wave 
function correction, which is large for $\sq=\st$. 

The renormalization of $A_b$ requires special attention:  
For a given set of ($m_{\sb_1}$, $m_{\sb_2}$, $\t_{\sb}$, 
$M_b$), the value of the running $A_b$ can be 
numerically very different from the on--shell $A_b$, especially 
for large $\tan\beta$. 
In Eq.~(\ref{e25}) there are two sources for a large difference $\delta A_b$.  
One is the correction due to large $\delta m_b$. 
The other is the correction to the first term of Eq.~(\ref{e25}), which is 
the left-right mixing mass of $\sb$, $M_{LR}^2=m_b(A_b-\mu\tan\beta)$, 
at the tree--level. The QCD correction to this term is of the 
order of $\alpha_s M_{LR}^2$ and gives a contribution 
$\delta A_b={\cal O}(\alpha_s\mu\tan\beta)$, which is very large for 
large $\tan\beta$. 
The on--shell $A_b$ is therefore physically inconvenient, at least for
the large $\tan\beta$ case. In fact, this is due to the property 
that $\t_{\sb}$ is insensitive to $A_b$. 

Instead of on--shell $A_b$, we can use the on--shell $\theta_{\sb}$ as 
an input parameter. However, in order to avoid fine tuning of the 
parameters so that we do not get a too large value of running $A_b$, 
we rather use $A_b(Q=m_\phi)$ as input. 

\section{Numerical analysis}

In this section we show the full one--loop
SUSY QCD corrected widths of the MSSM Higgs 
boson decays into bottom quarks and squarks (including real gluon emission 
to avoid infrared divergence), namely $(h^0, H^0, A^0)\rightarrow b\bar{b}$, 
$H^+\rightarrow t\bar{b}$, $(H^0,A^0)\rightarrow\sb_i\sb_j^*$, and 
$H^+\rightarrow\st_i\sb_j^*$ ($i,j=1,2$), with and without the 
improvement worked out in this paper. 

\subsection{Procedure of the numerical calculation}

Our input parameters are all on--shell except $A_b$ which is
running as explained above, i.~e. we have
$M_t, M_b, \msQ(\st), \msU, \msD, A_t, A_b(Q), \mu$, $\tan\beta, m_A$, and 
$m_\sg$, with $Q$ at the mass of the decaying Higgs boson.
(Since real gluinos do not appear in our processes we take $m_\sg$ as a
tree--level input, neglecting radiative 
corrections \cite{dmsg} to $m_\sg$.)
For the kinematics (phase space) the on--shell masses are
used. For consistency, all arguments in the Passarino--Veltman
integrals, which appear in the calculation of the
one--loop corrections, are also taken on--shell. 
According to the procedure of $\overline{\rm DR}$ renormalization 
we set the UV divergence parameter 
$\Delta\,(= \frac{2}{\epsilon} - \gamma + \log4\pi)$ to zero.    
In the following we will describe in detail the procedure 
for obtaining all necessary on--shell and $\overline{\rm DR}$ 
parameters for quarks and squarks, namely running $m_q$, 
pole $m_{\sq_{1,2}}$, running $(M_{\sQ},\,M_{\sU},\,M_{\sD},\,A_t)$, 
and running and on--shell $\t_{\sq}$. 
 
{\bf Top--stop sector}: The on--shell masses $m_{\st_1}$, $m_{\st_2}$,
and the mixing angle $\theta_\st$ are calculated by
diagonalizing the well--known stop mass matrix in the $\st_L$--$\st_R$ 
basis. Using Eqs.~(\ref{e3}) and (\ref{e4}) we get $m_t(Q)_{\rm SM}$,
and using Eq.~(\ref{e23})
$m_t(Q) \equiv m_t(Q)_{\rm MSSM} = m_t(Q)_{\rm SM} + \delta
m_t^{(\sg)}$.
The counterterms $\delta m_{\st_i}^2$ follow from
\begin{equation} 
\delta m_{\sq_i}^2  =  \mbox{Re}[\Pi_{ii}^{\sq\,(g)}(m_{\sq_i}^2) +
\Pi_{ii}^{\sq\,(\sg)}(m_{\sq_i}^2) + \Pi_{ii}^{\sq\,(\sq)}]\, ,
\label{dmsq}
\end{equation}
using Eqs.~(25)--(27) of \cite{hsq}.
For $\delta\theta_\st$ we take Eq.~(\ref{e24}) with $\Pi^\sq_{12}(p^2)$ as
given in \cite{hsq}. 
Note that the self--energies $\Pi_{ij}$ depend on the 
$\overline{{\rm DR}}$ scale $Q$.
Now we can compute the running stop masses 
$m_{\st_i}^2(Q) = m_{\st_i}^2 + \delta m_{\st_i}^2$ and 
$\theta_\st(Q) = \theta_\st + \delta\theta_\st$. Inserting the 
running masses and mixing angle into the formulas
\begin{eqnarray}
M_{\sQ}^{2} & = & m_{\st_1}^2 \cos^{2}\theta_{\st} +
m_{\st_2}^2 \sin^{2}\theta_{\st} - m_{t}^{2} - D_{L}(\st)\, ,
 \label{eq:mQ1}
                        \\
M_{\sU}^{2} & = & m_{\st_1}^2 \sin^{2}\theta_{\st} +
m_{\st_2}^2 \cos^{2}\theta_{\st} - m_{t}^{2} - D_{R}(\st) \, ,
 \label{eq:mU1}
                        \\
m_t\,A_t & = & (m_{\st_1}^2 - m_{\st_2}^2)\, \sin\theta_\st \,\cos\theta_\st 
+ m_t \,\mu \,\cot\beta\, ,  
  \label{eq:aqrun}
\end{eqnarray} 
with the $D$--terms
\begin{equation}
D_{L}(\sq) = m_Z^2\cos 2\beta (I^{3L}_q - e_q\sin^2\theta_W)\, , 
\qquad
D_{R}(\sq) = m_Z^2\cos 2\beta e_q\sin^2 \theta_W\,,
\end{equation}
we finally get the running parameters $M_{\sQ}(Q)$, $M_{\sU}(Q)$,
and $A_t(Q)$. $A_t(Q)$ will be needed for calculating the
Higgs--stop--stop and Higgs--stop--sbottom couplings.

{\bf Bottom--sbottom sector}: 
Here the situation is more complicated because the input parameter
$A_b$ is the running value at $Q = m_\phi$ and all other parameters are
on--shell. Therefore, we have to perform an iteration procedure to obtain 
$m_b(Q)_{\rm MSSM}$ and on--shell $m_{\sb_{1,2}}$ and $\theta_{\sb}$.\\

First we calculate the starting values for this iteration, which we
denote by a superscript $(0)$.
From the stop sector we already know the running parameter
$M_{\sQ}(Q)$. From Eqs.~(\ref{e3}) and (\ref{e4}) we obtain
$m_{b}(Q)_{\rm SM}$.
Together with the on--shell $M_{\sD}$, $\tan\beta$, and $\mu$
we calculate $m_{\sb_{1,2}}^{(0)}$ and $\theta_\sb^{(0)}$ by solving the
corresponding mass eigenvalue problem, see Eqs. (1) to (5) of \cite{hsq}. 
Next we evaluate the gluino
correction term $\delta m_b^{(\sg, 0)}$ using $m_\sg$,
$m_{b\, {\rm SM}}$, $m_{\sb_{1,2}}^{(0)}$, and $\theta_\sb^{(0)}$ and
then $m_b^{(0)} = m_{b\, {\rm SM}} + \delta m_b^{(\sg, 0)}$.
As the value of the running $M_{\sD}$ is similar to its on--shell
value we set $M_{\sD}^{(0)} = M_{\sD}$.
The values $\tan\beta$ and $\mu$ are not
affected by QCD and remain constant. The running value 
$M_{\sQ}$ will not be iterated because it is already calculated precisely
enough from the stop sector.\\ 

\newpage
\noindent The iteration procedure is as follows:      
\begin{enumerate}
\item $m_{\sb_{1,2}}^{(n)}$ and $\theta_{\sb}^{(n)}$ are
calculated
from the parameters $M_{\sQ}(Q)$, $A_b(Q)$, $M_{\sD}^{(n-1)}$, and
$m_b^{(n-1)}$.

\item We calculate 
$\delta m_b^{(\sg, n)}$ according to Eq.~(\ref{e23}) using 
$m_{\sb_{1,2}}^{(n)}$, $\theta_{\sb}^{(n)}$, and $m_b^{(n-1)}$
(instead of $M_b$).

\item 
$m_b^{(n)} = {m_b}_{\rm SM} + \delta m_b^{(\sg, n)}$.
 
\item $\delta m_{\sb_{1,2}}^{(n)\, 2}$ and 
$\delta \theta_{\sb}^{(n)}$ are calculated 
from $m_{\sb_{1,2}}^{(n)}$, $\theta_\sb^{(n)}$, and
$m_b^{(n)}$ using Eqs.~(\ref{dmsq}) and (\ref{e24}).

\item The on--shell values 
$m_{\sb_{1,2}\,{\rm os}}^{(n)} = 
\sqrt{m_{\sb_{1,2}}^{(n)\,2} - \delta m_{\sb_{1,2}}^{(n)\,2}}$, and
$\theta_{\sb\,{\rm os}}^{(n)} = 
\theta_{\sb}^{(n)} - \delta \theta_{\sb}^{(n)}$.

\item
$\delta M_{\sD}^{(n)\, 2}  = \delta m_{\sb_1}^{(n)\, 2}
\sin^{2}\theta_{\sb\,{\rm os}}^{(n)} +
\delta m_{\sb_2}^{(n)\,2} \cos^{2}\theta_{\sb\,{\rm os}}^{(n)} 
+ (m_{\sb_1}^2 - m_{\sb_2}^2)_{\rm os}^{(n)}
\sin 2\theta_{\sb\,{\rm os}}^{(n)} 
\delta\theta_{\sb}^{(n)}
- 2 M_{b} \delta m_{b}^{(n)}\, ,$ with
$\delta m_{b}^{(n)} = m_{b}^{(n)} - M_b$. 

\item $M_{\sD}^{(n)} = \sqrt{M_{\sD}^{2} + \delta M_{\sD}^{(n)\,2}}$.

\end{enumerate}

\noi
Note that all quantities $X^{(n)}$ without the subscript ``OS'' 
are $\overline{\rm DR}$ running ones.

We define an accuracy parameter $\Delta (x)$ by
$\Delta (x) = \Big| 1 - \frac{x^{(n)}(Q)}{x^{(n-1)}(Q)} \Big|$.
The iteration starts with $n=1$ 
and stops, when $\Delta (M_{\sD})$,
$\Delta (m_b)$, and $\Delta (\theta_{\sb})$ are all smaller than 
a given accuracy. In our analysis we require $\Delta (x) < 10^{-5}$.
Thus we obtain $m_{\sb_{1,2}\,{\rm os}}$, $\theta_{\sb\,{\rm os}}$, 
and $m_b(Q)$, which we need 
for the calculation of the Higgs boson decay widths. 
The consistency of this procedure is checked 
by calculating the on--shell
$M_{\sD}$ from
\begin{equation}
M_{\sD}^{2} =  m_{\sb_1\,{\rm os}}^2 \sin^{2}\theta_{\sb\,{\rm os}} +
m_{\sb_2\,{\rm os}}^2 \cos^{2}\theta_{\sb\,{\rm os}} - 
M_{b}^{2} - D_{R}(\sb) \, ,
 \label{eq:mD}  
\end{equation}
which must be equal to the input $M_{\sD}$ within the same order
of the above accuracy.\\
  
{\bf Decays into quarks}: We first consider the tree--level as the
zeroth approximation. As described in section~\ref{sec3.2}, for the
decays of $A^0$ and $H^\pm$ we take the
running mass $m_{t,b}(m_\phi) \equiv m_{t,b}(m_\phi)_{\rm MSSM}$ 
in the Yukawa couplings, i.e.
\begin{eqnarray}
&& y_{1} = -i \sqrt{2}\, a^{b} = h_b \sin\beta =
   \sqrt{2} \,\frac{m_{b}(m_{A})}{\bar v}\, \tan\beta \, ,\nonumber \\
&& y_{2} = -i \sqrt{2}\, a^{t} = h_t \cos\beta = 
   \sqrt{2} \,\frac{m_{t}(m_{A})}{\bar v}\,\cot\beta\, ,  
\label{e37}
\end{eqnarray}
with $\bar v = 2 m_{W}/g$ and 
$y_{1,2}$ the $H^+\bar tb$ couplings, see Eq.~(20) of \cite{hsq}.  
As outlined in section~\ref{sec3.2}, for the
couplings of $h^0$ and $H^0$ to top and bottom quarks 
we use
\begin{eqnarray}
 s_{h}^{t} & = & -  \frac{m_{t}(m_{h})}{\bar v}\, 
\frac{\cos\alpha}{\sin\beta}\, ,
                        \nonumber\\
 s_{H}^{t} & = & -  \frac{m_{t}(m_{H})}{\bar v}\, 
\frac{\sin\alpha}{\sin\beta}\, , 
                        \nonumber\\
s_{h}^{b}  & = &  \sin(\alpha-\beta) \frac{m_{b}(m_{h})_{\rm SM}}{\bar v} +
            \cos(\alpha - \beta)\, \tan\beta\, \frac{m_{b}(m_{h})}{\bar v}\, ,
                        \nonumber\\
s_{H}^{b}  & = & -\cos(\alpha-\beta) \frac{m_{b}(m_{H})_{\rm SM}}{\bar v} +
           \sin(\alpha - \beta)\, \tan\beta\, \frac{m_{b}(m_{H})}{\bar v}\, .
  \label{e38}
\end{eqnarray}
One has to bear in mind that in the above
procedure we have already absorbed the counterterms to $m_q$ with gluon and
gluino exchange in the running quark mass $m_q(Q)$ in 
$a^t$, $a^b$, $s_h^t$, and $s_H^t$. Therefore the corresponding
counterterms $\delta a^{t\,(0)}$, $\delta a^{b\,(0)}$, 
$\delta s_{h}^{t\,(0)}$, and $\delta s_{H}^{t\,(0)}$ are 
zero\footnote{Here the superscript $(0)$ denotes the counterterms to
the couplings; notation as in \cite{hsq}.}. Only
for the decays  of $h^0$ and $H^0$ into bottom quarks the gluino
contribution is just partly included. Therefore the counterterms are
\begin{eqnarray}
\delta s_{h}^{b\,(0)} & = & \sin(\alpha-\beta)\, 
\frac{\delta m_{b}^{(\sg)}}{\bar v}\, , 
                       \nonumber\\
\delta s_{H}^{b\,(0)} & = & -\cos(\alpha-\beta)\, 
\frac{\delta m_{b}^{(\sg)}}{\bar v}\, . 
\end{eqnarray}
In the Higgs--squark--squark couplings $G_{ijk}^{\sq}$ in the loop 
(for notation, see \cite{hsq}) $m_q$, $\theta_\sq$, and $A_q$ are all taken
running at the scale $m_\phi$. 

{\bf Decays into squarks}:
The tree--level matrix element is directly proportional to the 
Higgs--squark--squark couplings $G_{ijk}^{\sq}$
which are functions of $m_q, A_q$, and $\theta_\sq$.
According to section~\ref{sec4} we use running $m_q$ and $A_q$,
but on--shell $\theta_\sq$. As we have absorbed the counterterms for
$m_{q}$ and
$A_q$ in the tree--level couplings, we have the shifts $\delta m_{q} =
\delta A_q = 0$ in the calculation of the one--loop corrections. 
The remaining counterterms are 
($i'\ne i$ and $j' \ne j$)
\begin{equation}
        \label{eq:dGij0}
        \delta G_{ijk}^{\sq\,(0)} =
        - \left((-1)^i G_{i'jk}^\sq + (-1)^j G_{ij'k}^\sq\right)
        \delta \theta_\sq 
\end{equation}
for $H_k = \{ h^0, H^0 \}$. For $H^+ \to \st_i \sb_j^*$ we have
\begin{equation}
\label{eq:dGij40}
\delta G_{ij4}^{(0)} = 
 - (-1)^i G_{i'j4}\, \delta \theta_\st -
 (-1)^j G_{ij'4} \,\delta \theta_\sb \, .
\end{equation}
The tree--level coupling $A^0 \sq_1^* \sq_2$ is independent of
$\theta_\sq$. Therefore no counterterms are left in this case. 
For the Higgs couplings to quarks in the loop we take Eqs.~(\ref{e37})
and (\ref{e38}).\\
 
In the {\bf numerical analysis}
we take for the Standard Model parameters $M_t=175$ GeV, 
$M_b=5$ GeV, $\a_s(m_Z)=0.12$, $m_Z=91.2$ GeV, 
$\sin^2\t_W = 0.23$, and $\a_{em}(m_Z) = 1/129$. 

Although we treat only the SUSY QCD corrections, the inclusion of the 
very large Yukawa correction to the Higgs boson sector \cite{higgscorr}
is indispensable for a realistic study. We thus use the formulae of 
ref.~\cite{carena} for the radiative corrections to ($m_{h^0}$, $m_{H^0}$, 
$\alpha$). For $m_{H^+}$ the tree--level formula is used, which is a
very good approximation for $m_A \gg m_Z$.

We choose the Higgs and SUSY parameters such that they
satisfy the mass bounds from 
direct searches \cite{search}, 
the constraint from electroweak $\delta\rho$ bounds on 
$\st$ and $\sb$ \cite{deltarho} using the formula of \cite{dreeshag}, 
and the approximate necessary 
condition for the tree--level vacuum stability \cite{stability}, 
\beq  \label{e27}
A_t^2(Q) < 3\,(M_{\ti Q}^2(Q) + M_{\ti U}^2(Q) + m_{H_2}^2),\;\;\;
A_b^2(Q) < 3\,(M_{\ti Q}^2(Q) + M_{\ti D}^2(Q) + m_{H_1}^2)
\eeq
with $m_{H_2}^2=(m_{A}^2+m_{Z}^2)\cos^2\b-\frac{1}{2}\,m_Z^2$,  
$m_{H_1}^2=(m_{A}^2+m_{Z}^2)\sin^2\b-\frac{1}{2}\,m_Z^2$, 
and $Q \sim M_{\tilde Q}$. 

\subsection{Numerical results and discussion}
\begin{figure}[htbp]
  \begin{center}
\mbox{\epsfig{figure=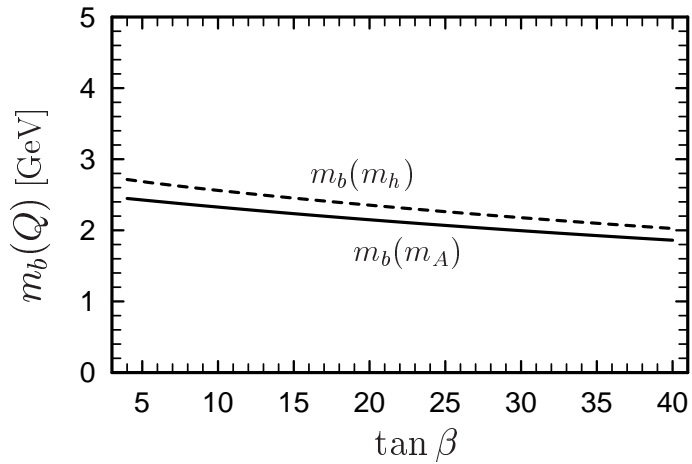,width=10cm}}
  \end{center}
\vspace{-5mm}
\caption[fig1]{Running bottom mass as a function of $\tan\beta$.
The on--shell value is
$M_b = 5$~GeV. Furthermore, $m_b(m_h)_{\rm SM} \simeq 3.0$~GeV with
$m_{h^0} = 93$ to $103$~GeV, and
$m_b(m_A)_{\rm SM} = 2.6$~GeV. 
The other parameters are given in the text. \label{fig:1}} 
\end{figure}

In the following we show the numerical improvement of the SUSY QCD corrections 
to the widths of Higgs boson decays to quarks and squarks.
First we show in Fig.~\ref{fig:1} the SUSY QCD running bottom quark mass
at the scale of $h^0$, $m_b(m_h)$ (dashed line),
and at the scale of $A^0$, $m_b(m_A)$ (full line), as a function of 
$\tan\beta$. 
The following values for the parameters are used:
$(M_{\tilde{Q}}(\st),M_{\tilde{U}},
M_{\tilde{D}})=(300,270,330)$ GeV, 
$A_t=150$ GeV, $A_b(Q)=-700$ GeV (where $Q=m_h$ or $m_A$), 
$m_{\sg}=350$~GeV,
$\mu=260$~GeV, and $m_A=800$~GeV. 

\begin{figure}[htbp]
  \begin{center}
\hspace{2mm} \mbox{\epsfig{figure=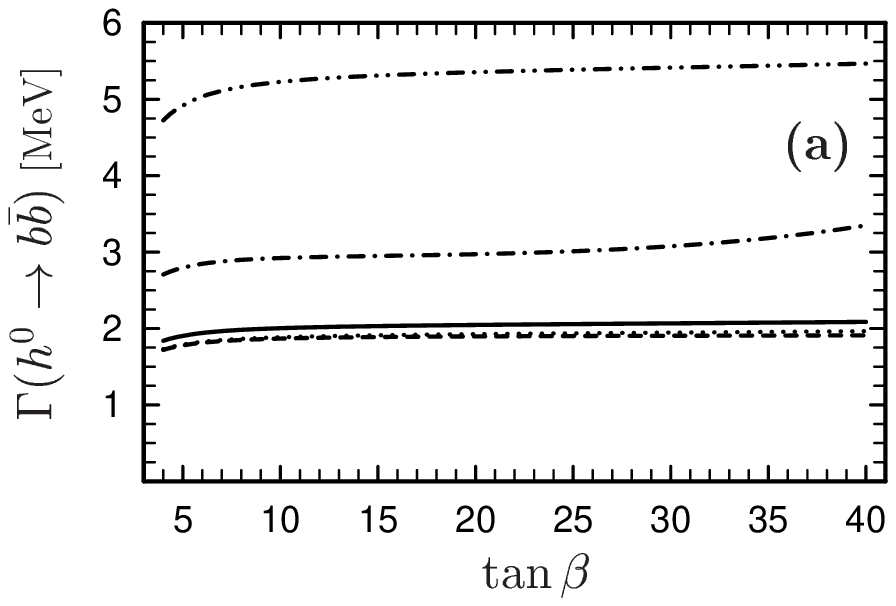,width=10.2cm}}\\
\vspace{0.4cm}
\mbox{\epsfig{figure=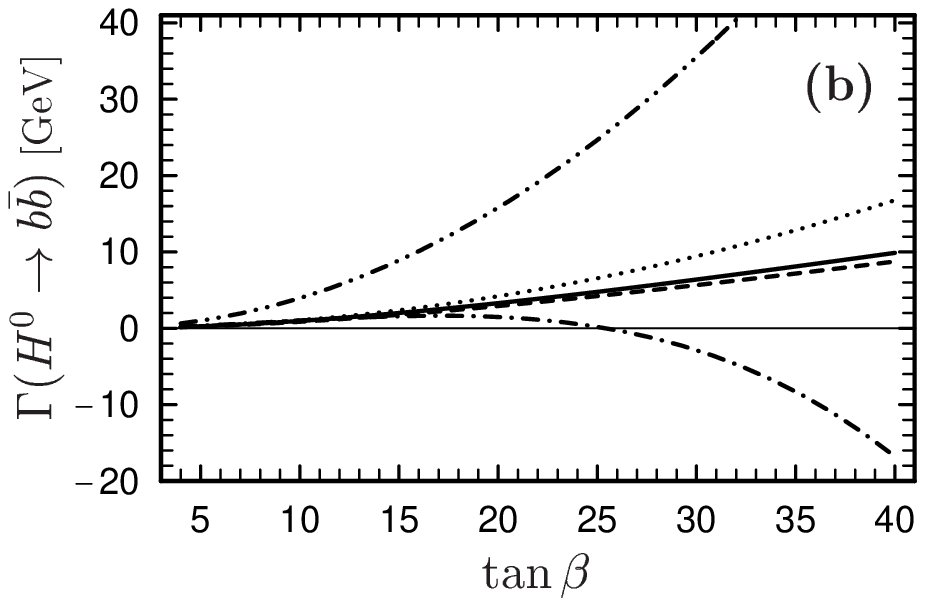,width=10cm}}\\
\vspace{0.4cm}
\mbox{\epsfig{figure=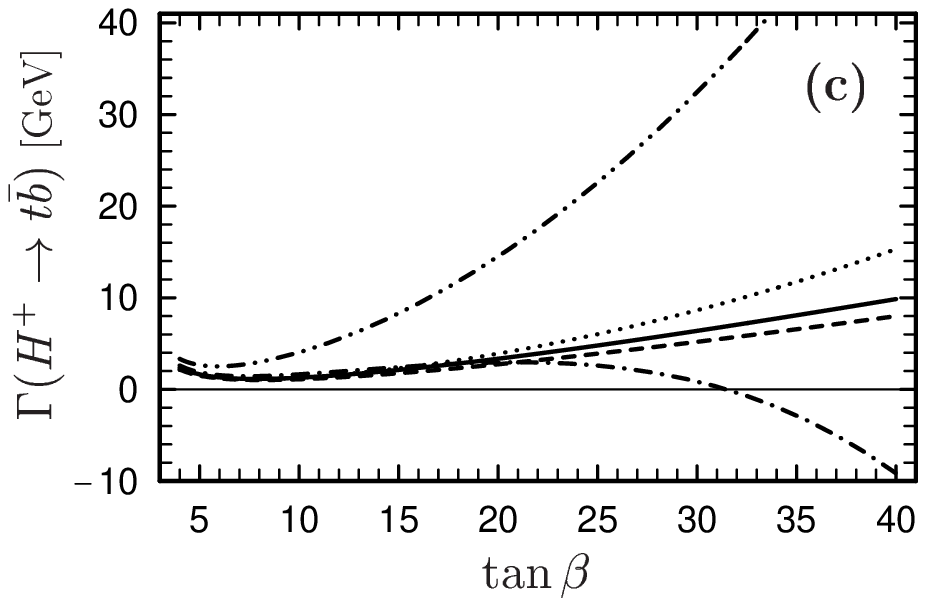,width=10cm}}
  \end{center}
\vspace{-5mm}
\caption[fig1a]{Widths of Higgs particle decays into quarks as a
function of $\tan\beta$.
Case (i): dash--dot--dotted line corresponds to the on--shell tree--level,
and dash--dotted to the on--shell one--loop result.
Case (ii): dashed line corresponds to the improved tree--level,
and full line to the improved one--loop result.
Case (iii): dotted line corresponds to the tree--level result improved
only by using $m_b(Q)_{\rm SM}$.
For details, see the related text.   
\label{fig:1a}} 
\end{figure}

In Fig.~\ref{fig:1a} the tree--level
and corrected widths of Higgs boson decays to bottom 
quarks are shown in two ways of perturbative expansion, 
(i) the strict on--shell scheme (dash--dot--dotted line: tree--level, 
dash--dotted line: one--loop corrected), 
and (ii) the improved scheme as
discussed in section~\ref{sec3.2} (dashed: tree--level,
full line: one--loop corrected), as a function of
$\tan\beta$. The parameters used are the same as in Fig.~\ref{fig:1}
with $Q$ at the mass of the parent Higgs boson.  
(Notice that for case~(i) $A_b=-700$~GeV is the on--shell value.)
In addition, the tree--level decay width using
(iii) the non--supersymmetric QCD running mass $m_b(m_\phi)_{\rm SM}$ is
shown (dotted line).
One can clearly see that the difference between tree--level and corrected 
widths decreases dramatically from (i) to (ii), 
especially for decays of the heavier Higgs bosons $(H^0, A^0, H^\pm)$. 
(The curves for $\Gamma(A^0\to b\bar b)$ are practically the same 
as those for $\Gamma(H^0\to b\bar b)$.)
In particular, the physically meaningless ``negative width'' 
is absent for (ii). This demonstrates the improvement of 
the convergence of the SUSY QCD perturbation expansion by the method 
proposed in this paper. Furthermore, Fig.~\ref{fig:1a}a shows 
that (iii) is indeed a good approximation for $h^0 \to b \bar b$ in the 
large $m_A$ limit, Eq.~(\ref{e17}). This 
is, however,
not the case for the decays of the heavier Higgs particles to quarks,
see the Figs.~\ref{fig:1a}b and \ref{fig:1a}c.\\ 

\begin{figure}[htbp]
  \begin{center}
\hspace{3mm} \mbox{\epsfig{figure=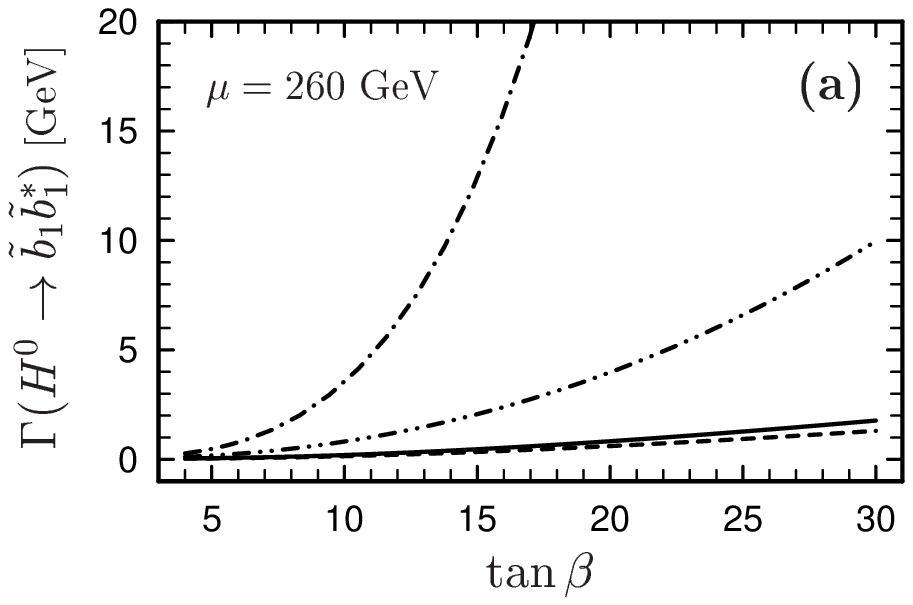,width=10.3cm}}\\
\vspace{0.4cm}
\hspace{2mm} \mbox{\epsfig{figure=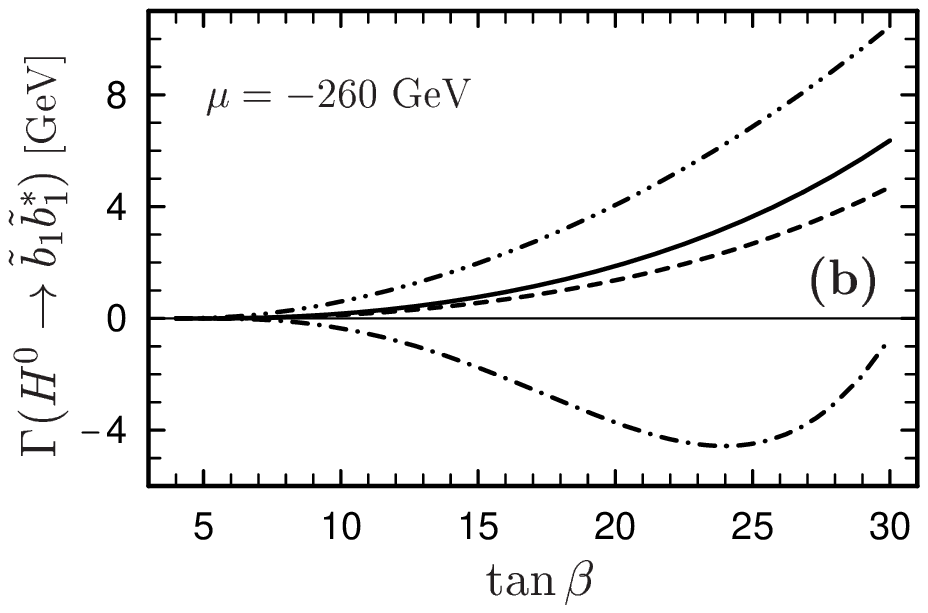,width=10cm}}\\
\vspace{0.4cm}
\hspace{3mm} \mbox{\epsfig{figure=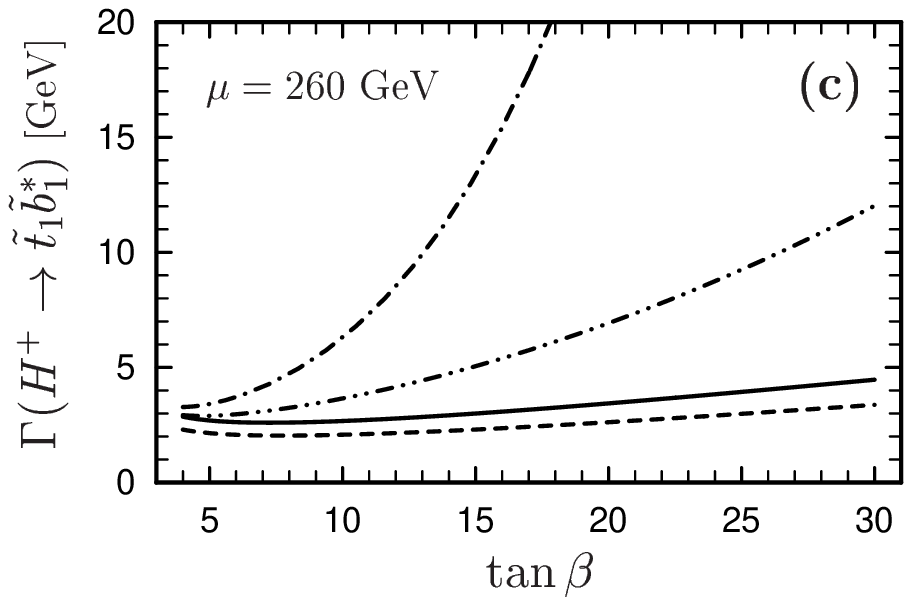,width=10.3cm}}
  \end{center}
\vspace{-5mm}
\caption[fig2]{Widths of Higgs particle decays into squarks as a
function of $\tan\beta$. 
Case (i): dash--dot--dotted line corresponds to the on--shell tree--level,
and dash--dotted to the on--shell one--loop result.
Case (ii): dashed line corresponds to the improved tree--level,
and full line to the improved one--loop result.
For details, see the related text. \label{fig:2}} 
\end{figure}

We next show the numerical improvement of the SUSY QCD corrections 
to the decay widths to squarks. 
Figure~\ref{fig:2} shows the tree--level and corrected 
decay widths of 
$H^0\rightarrow\sb_1\sb_1^*$ for $\mu = 260$~GeV and 
$-260$~GeV, and
$H^+\rightarrow\st_1^{}\sb_1^*$ for $\mu = 260$~GeV,  
as functions of $\tan\beta$. 
The other parameters are the same as in Fig.~\ref{fig:1a}.  
We compare the strict on--shell perturbation
expansion of \cite{hsq} (i) and the improved one (ii) as described 
in section~\ref{sec4}.
The improvement of the perturbation expansion is again very clear. 
The results for $\Gamma(A^0\to \sb_1\sb_2^*)$ are similar 
to those for $\Gamma(H^0\to \sb_1\sb_1^*)$.

\begin{figure}[htbp]
  \begin{center}
\mbox{\epsfig{figure=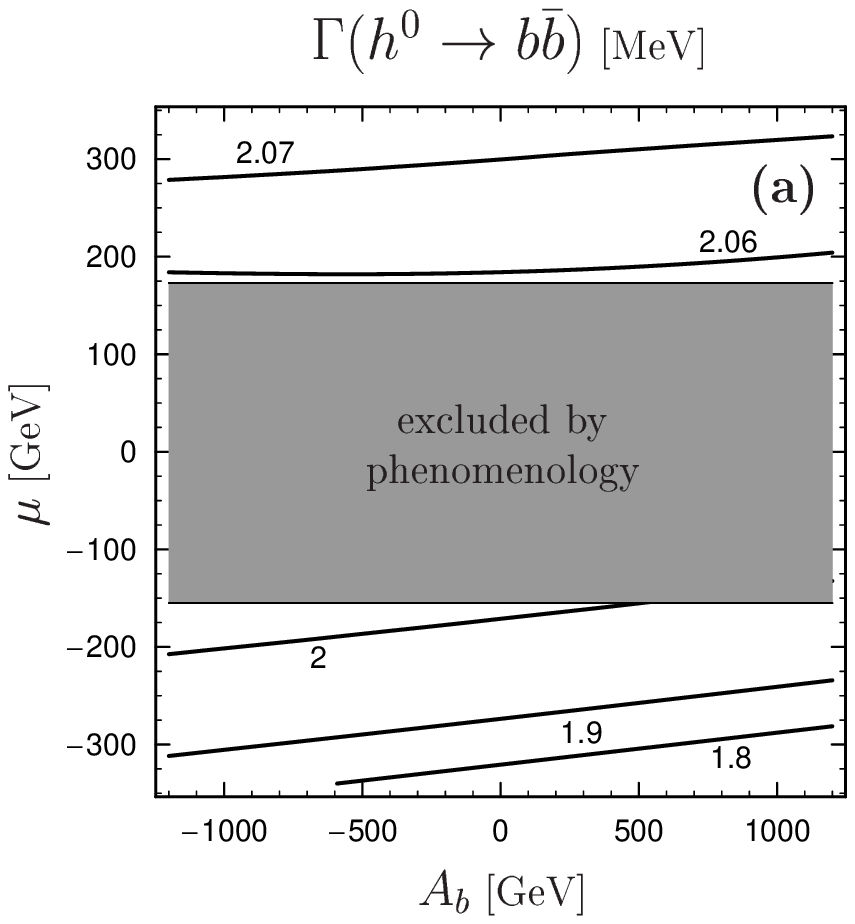,width=10cm}}
\vspace{1cm}
\mbox{\epsfig{figure=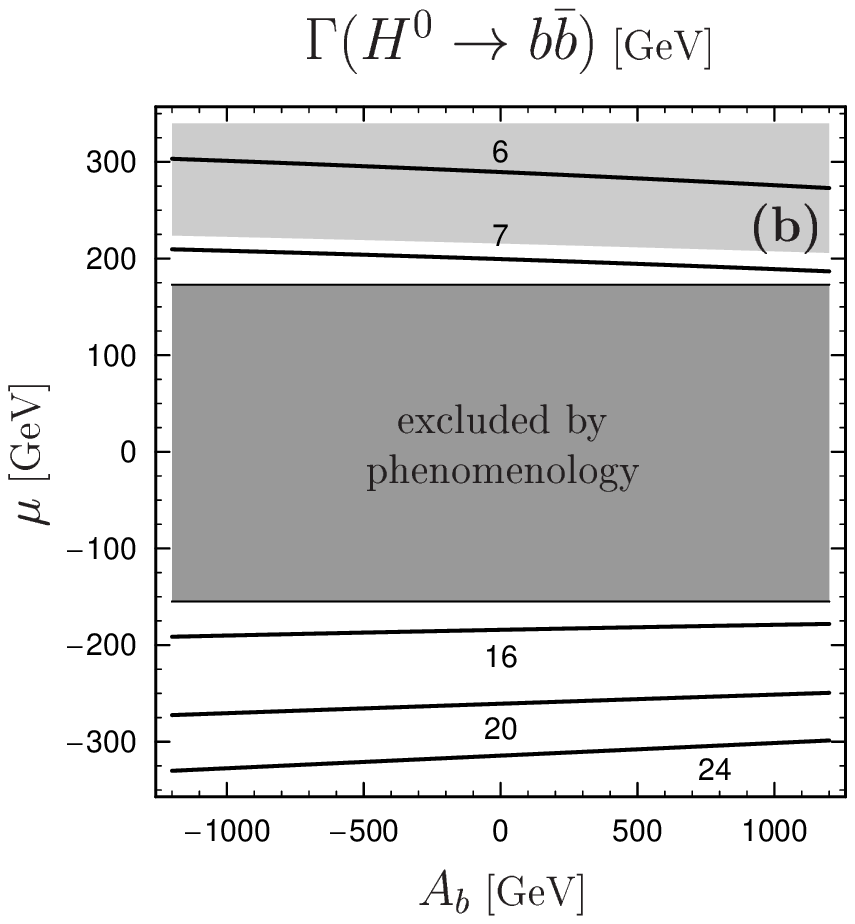,width=10cm}}
  \end{center}
\vspace{-16mm}
\caption[fig3]{Contour lines for the improved one--loop decay
widths of
Higgs particle decays into bottom quarks as a function of $A_b$ running and
$\mu$. In (b) the light gray area shows the region where the on--shell
one--loop result is negative. The other parameters are given in
the text. The dark gray area is excluded by 
LEP and Tevatron data \cite{search}. 
\label{fig:3}} 
\end{figure}

\begin{figure}[htbp]
  \begin{center}
\mbox{\epsfig{figure=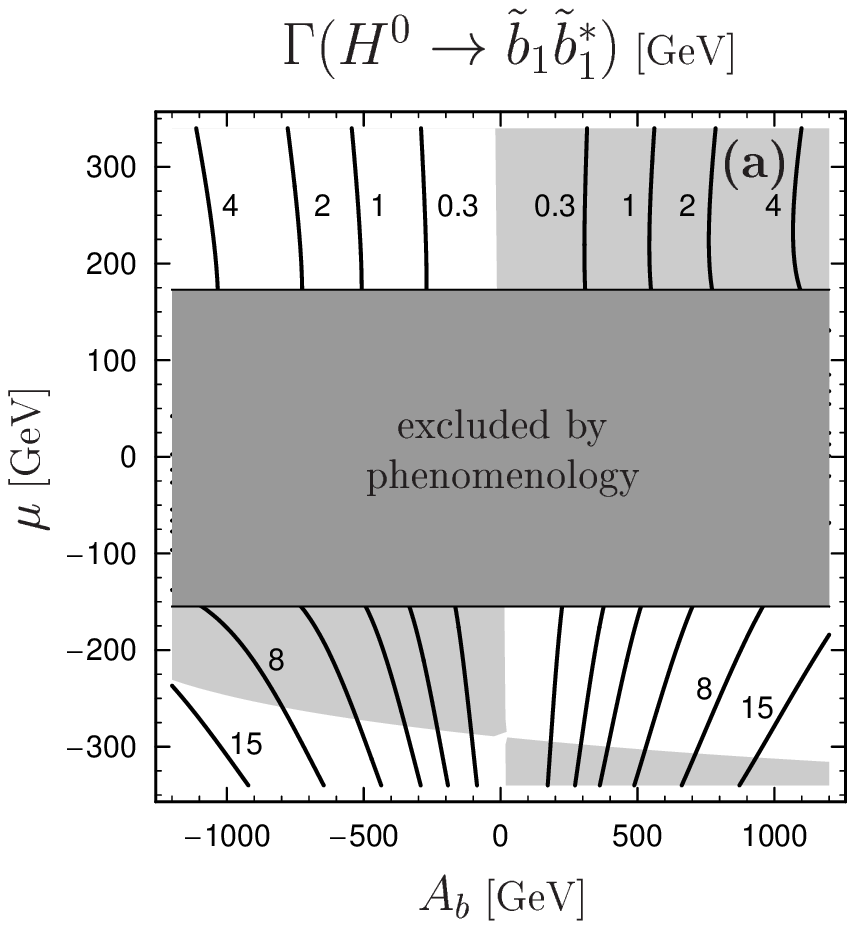}}
\vspace{1cm}
\mbox{\epsfig{figure=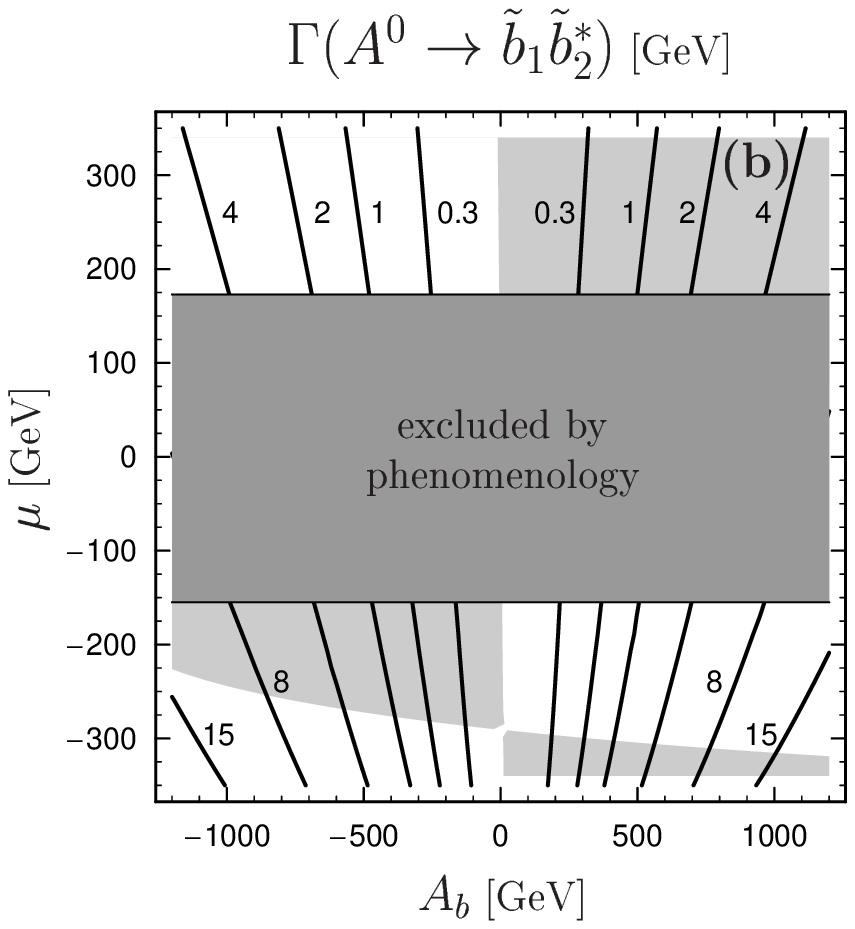}}
  \end{center}
\vspace{-13mm}
\caption[fig4]{Contour lines for the improved one--loop decay
widths of
Higgs particles decay into bottom squarks as a function of $A_b$ running and
$\mu$. The light gray areas show the region where the on--shell
one--loop result is negative. The other parameters are given in
the text. The dark gray area is excluded by 
LEP and Tevatron data \cite{search}. 
\label{fig:4}} 
\end{figure}

We now calculate the parameter dependence of the SUSY QCD corrected widths 
$\Gamma(h^0\rightarrow b\bar{b})$,
$\Gamma(H^0\rightarrow b\bar{b})$, $\Gamma(H^0\rightarrow \sb_1\sb_1^*)$,
and $\Gamma(A^0\rightarrow\sb_1^{}\sb_2^*)$ 
for $\tan\beta=30$, which could not be presented in \cite{htb,hsq} 
because the widths would have become negative. 
The results are shown in Figs.~\ref{fig:3} and \ref{fig:4} as 
contour plots in the ($A_b(Q)$, $\mu$) plane for $\tan\beta=30$, 
with the SU(2) gaugino mass $M = 117$~GeV, and the
other parameters being the same as in Fig.~\ref{fig:1}. 

As expected from Eq.~(\ref{e8}), in Fig.~\ref{fig:3} the corrections to 
$\Gamma(h^0\rightarrow b\bar{b})$ and
$\Gamma(H^0\rightarrow b\bar{b})$ are mainly determined by $\mu$.
Note also that the $\mu$ dependence is much stronger for 
$\Gamma(H^0\rightarrow b\bar{b})$ in Fig.~\ref{fig:3}b than for 
$\Gamma(h^0\rightarrow b\bar{b})$ in Fig.~\ref{fig:3}a, in accordance 
with the discussion in section 3.1. 

As can be seen in Fig.~\ref{fig:4}a, the decay width 
of $H^0\rightarrow\sb_1\sb_1^*$ becomes large for 
large $|A_b(Q)|$ and/or 
large negative $\mu$. One can see that the tree--level invariance of the 
decay width under the sign change $(A_b,\mu)\rightarrow(-A_b,-\mu)$ 
is badly violated. This is due to the gluino loops. 

Finally we comment on the $m_{\sg}$ dependence. 
It has been pointed out in \cite{hbb,htb2}
that the gluino correction to quark modes 
decouples very 
slowly for increasing $m_{\sg}$ $(m_{\sg} > m_{\sq})$
with fixed squark parameters. 
This property can be understood from Eq.~(\ref{e8}). 
The correction $\Delta_b$ scales as $(\ln m_{\sg})/m_{\sg}$ in this 
limit, decreasing much more slowly than the usual $1/M^2$ behavior. 

\section{Conclusion}

The SUSY QCD corrections to the decays of MSSM Higgs 
bosons to quarks and squarks of the third gereration are 
phenomenologically very important. 
The existing formulae for the one--loop SUSY QCD corrections 
to their decay widths have been given in the on--shell expansion for 
quarks and squarks. However, for large $\tan\beta$, these formulae 
suffer from very bad convergence of the perturbation series, which makes 
the numerical results unreliable. 

We have worked out a method for improving the SUSY QCD corrections to 
these widths. 
The essential point of the improvement is to define appropriate 
tree--level couplings of the Higgs bosons to quarks and 
squarks, in terms of the running quark masses $m_q(Q)$ both in 
non--SUSY and SUSY QCD, the running Higgs--squark trilinear 
couplings $A_q(Q)$, 
and the on--shell left--right mixing angle $\theta_{\sq}$ of squarks. 
We have also shown numerically that this method greatly 
improves the corrections to the decays to bottom quarks and squarks. 

We finally note that the method of improvement for the Higgs boson couplings 
to quarks and squarks as presented in this paper is also useful in studying 
radiative corrections to other processes, 
such as the decays of a squark into 
chargino/neutralino and quark \cite{sqchnt,beenakker3}, 
those into lighter squark and Higgs boson \cite{hsq2,sqhx}, 
low energy four--fermi interactions with virtual charged 
Higgs boson \cite{hpvirtual}, 
and associated production of Higgs bosons with quarks \cite{qqhass} and 
squarks \cite{eesqsqh}. 

\section*{Acknowledgements}

The work of Y.\,Y. was supported in part by the Grant--in--aid for Scientific 
Research from the Ministry of Education, Science, Sports, and Culture of 
Japan, No.~10740106. H.\,E. S.\,K., and W.\,M. thank the 
''Fonds zur F\"orderung der wissenschaftlichen Forschung of Austria'', 
project no. P13139-PHY for financial support.
We thank M. Drees for pointing out a misprint in Figs. 2a and 4a.


\end{document}